\documentclass[aps,twocolumn,superscriptaddress,amsmath,amssymb]{revtex4-2}

\usepackage{braket,mathtools}
\usepackage[unicode=true]{hyperref}
\usepackage[T1]{fontenc}

\renewcommand{\tilde}{~}

\newcommand{\trace}[1]{\text{Tr}\left[#1\right]}
\newcommand{\proj}[1]{\Pi_{\scriptscriptstyle\text{#1}}}
\newcommand{\base}[1]{\mathcal{B}_{\scriptscriptstyle\text{#1}}}

\newcommand{\kluru}{\ket{L\uparrow,R\uparrow}}
\newcommand{\klurd}{\ket{L\uparrow,R\downarrow}}
\newcommand{\kldru}{\ket{L\downarrow,R\uparrow}}
\newcommand{\kldrd}{\ket{L\downarrow,R\downarrow}}

\newcommand{\bldrd}{\bra{L\downarrow,R\downarrow}}

\newcommand{\klulu}{\ket{L\uparrow,L\uparrow}}
\newcommand{\kluld}{\ket{L\uparrow,L\downarrow}}
\newcommand{\kldld}{\ket{L\downarrow,L\downarrow}}
\newcommand{\kruru}{\ket{R\uparrow,R\uparrow}}
\newcommand{\krurd}{\ket{R\uparrow,R\downarrow}}
\newcommand{\krdrd}{\ket{R\downarrow,R\downarrow}}

\newcommand{\kop}[1]{\ket{1_+}_{\scriptscriptstyle\textrm{#1}}}
\newcommand{\kom}[1]{\ket{1_-}_{\scriptscriptstyle\textrm{#1}}}
\newcommand{\ktp}[1]{\ket{2_+}_{\scriptscriptstyle\textrm{#1}}}
\newcommand{\ktm}[1]{\ket{2_-}_{\scriptscriptstyle\textrm{#1}}}

\newcommand{\kum}[1]{\ket{U_-}_{\scriptscriptstyle\textrm{#1}}}

\newcommand{\kdm}[1]{\ket{D_-}_{\scriptscriptstyle\textrm{#1}}}

\newcommand{\bum}[1]{\bra{U_-}_{\scriptscriptstyle\textrm{#1}}}

\newcommand{\bdm}[1]{\bra{D_-}_{\scriptscriptstyle\textrm{#1}}}
\newcommand{\bop}[1]{\bra{1_+}_{\scriptscriptstyle\textrm{#1}}}
\newcommand{\bom}[1]{\bra{1_-}_{\scriptscriptstyle\textrm{#1}}}
\newcommand{\btp}[1]{\bra{2_+}_{\scriptscriptstyle\textrm{#1}}}
\newcommand{\btm}[1]{\bra{2_-}_{\scriptscriptstyle\textrm{#1}}}

\newcommand{\rlr}{\rho_{\scriptscriptstyle{\text{LR}}}}

\newcommand{\rbs}{\rho_{\scriptscriptstyle\text{BS}}}

\newcommand{\rdep}{\rho_{\scriptscriptstyle{\text{dep}}}}

\newcommand{\rhogen}[2]{\rho_{\scriptscriptstyle\text{#1}}^{\scriptscriptstyle\text{#2}}}

\newcommand{\plr}{p_{\scriptscriptstyle\text{LR}}}
\newcommand{\pno}{p_{\scriptscriptstyle\text{NO}}}

\newcommand{\pgen}[2]{p_{\scriptscriptstyle\text{#1}}^{\scriptscriptstyle\text{#2}}}

\begin{document}

    \title{Robust engineering of maximally entangled states by identical particle interferometry}

    \author{Matteo Piccolini}
	\email{matteo.piccolini@unipa.it}
	\affiliation{Dipartimento di Ingegneria, Universit\`{a}      di Palermo, Viale delle Scienze, 90128 Palermo, Italy}
	
    \author{Vittorio Giovannetti}
        \affiliation{NEST, Scuola Normale Superiore and Istituto Nanoscienze-CNR, I-56126 Pisa, Italy}
	
    \author{Rosario Lo Franco}
	\email{rosario.lofranco@unipa.it}
	\affiliation{Dipartimento di Ingegneria, Universit\`{a} di Palermo, Viale delle Scienze, 90128 Palermo, Italy}

    \begin{abstract}
         We propose a procedure for the robust preparation of maximally entangled states of identical fermionic qubits, studying the role played by particle statistics in the process. The protocol exploits externally activated noisy channels to reset the system to a known state. The subsequent interference effects generated at a beam splitter result in a mixture of maximally entangled Bell states and NOON states. We also discuss how every maximally entangled state of two fermionic qubits distributed over two spatial modes can be obtained from one another by fermionic passive optical transformations. Using a pseudospin-insensitive, non-absorbing, parity check detector, the proposed technique is thus shown to deterministically prepare any arbitrary maximally entangled state of two identical fermions. These results extend recent findings related to bosonic qubits. 
Finally, we analyze the performance of the protocol for both bosons and fermions when the externally activated noisy channels are not used and the two qubits undergo standard types of noise.
The results supply further insights towards viable strategies for noise-protected entanglement exploitable in quantum-enhanced technologies.
    \end{abstract}

    \maketitle

    \section{Introduction}
    With the advent of technologies based on quantum paradigms, entanglement has become the subject of a rapidly increasing amount of studies\tilde\cite{horodecki}. These include, but are not limited to, its generation, manipulation, and protection from detrimental noise. The latter, in particular, is a crucial step to be tackled as quantum correlations are known to decay rapidly in systems exposed to the action of environmental noise\tilde\cite{aolita_2015}. Since achieving a perfectly isolated system at the quantum level is practically unfeasible, different strategies have been proposed over time to deal with entanglement fragility. These ranges from structured environments with memory effects\tilde\cite{mazzola_2009,bellomo_2008,lo_franco_2013,Xu_2010,bylicka_2014,man_2015,tan_2010,tong_2010,breuer_2016_colloquium,man_2015_pra}, decoherence-free subspaces\tilde\cite{zanardi1997noiseless, lidar1998decoherence}, dynamical decoupling and control techniques\tilde\cite{Viola1998,viola2005random,darrigo_2014_aop, franco2014preserving,orieux_2015,facchi_2004,lo_franco_2012_pra, xu_2013,damodarakurup_2009,cuevas_2017}, to quantum error corrections\tilde\cite{preskill_1998,knill2005quantum, shor_1995,steane_1996} and distillation protocols\tilde\cite{Bennett1996,horodecki1997inseparable,horodecki1998mixed,horodecki2001distillation,kwiat_2001,dong_2008}. Furthermore, a recently developed research line investigates the possibility to exploit the indistinguishability of identical particles to generate and protect quantum correlations in open quantum systems\tilde\cite{indistentanglprotection,Piccolini_2021_entropy,
    piccolini2022philtrans,piccolini2023asymptotically,nosrati2023indistinguishabilityassisted}. 
    
    A method has been recently proposed to distill pure maximally entangled states from two bosonic qubits subjected to arbitrary local noisy environments\tilde\cite{piccolini2023asymptotically}. To do so, this scheme exploits artificially induced noise to affect the two constituents after they have been subjected to the environmental noise sources. Allowing the externally activated noisy interaction to act for a sufficiently long time, the system is reset to a known state regardless of the environmental noise which previously affected it. Referring to a photonic implementation for simplicity, the resulting photons are later subjected to the non-local action of a beam splitter (BS). As a consequence of the interference effects occurring due to particle indistinguishability, the result is found to be a classical mixture of a Bell singlet state and NOON states. Using a polarization insensitive, non-absorbing, parity-check detector D set on one of the two spatial modes, the former is filtered probabilistically, while the latter are reintroduced in the BS and subjected once again to depolarization. In this way, the process can be iterated until a singlet is obtained, with distillation probability which scales to 1 exponentially. It has been also shown that all the maximally entangled states of two bosonic qubits distributed over two spatial modes can be mapped one into the other via \textit{passive optical} (PO) transformations and the action of the detector D\tilde\cite{piccolini2023asymptotically}, suggesting that the proposed procedure can be used to prepare any arbitrary maximally entangled state of two identical bosons. It is then important to study how a similar procedure can be applied to particles of different nature, such as fermions.    

    In this work, we first extend the procedure proposed in Ref.\tilde\cite{piccolini2023asymptotically} to identical fermions, introducing the analogous PO equivalences between fermionic maximally entangled states. After that, we analyze for both bosons and fermions the scenario where the artificially activated noise is not applicable to the two qubits. In this situation, where the system is not reset to a fixed state anymore, the performance of the scheme depends on the type of noisy environment affecting the particles, on the system-environment interaction time, and on the initially prepared state. In particular, we focus on three standard types of local noisy channels (phase damping, depolarizing, amplitude damping), and on a set of initial states which includes both entangled and separable ones.

    Multiparticle states are written in the \textit{no-label approach}\tilde\cite{nolabelappr,compagno2018dealing,slocc}, an alternative formalism for identical particles which allows to simplify the notation by avoiding to explicitly symmetrize/antisymmetrize global states as required by the symmetrization postulate in the standard first-quantization framework.

    \section{Procedure}
    
    \begin{figure}[t!]
        \includegraphics[width=0.9\columnwidth]{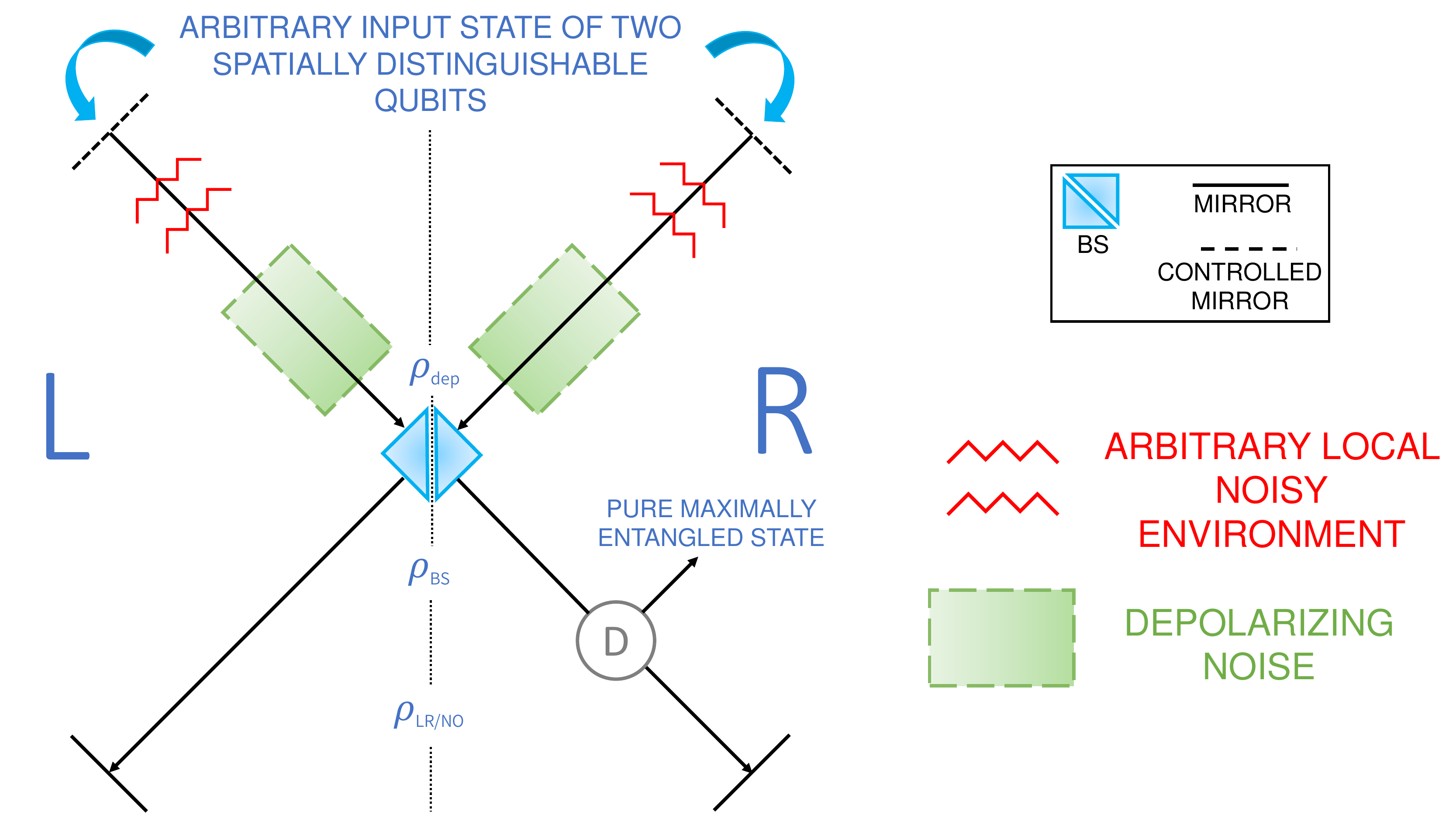}
        \centering
            \caption{\textbf{Schematic representation of the setup}. The scheme uses externally activated depolarization noises, here represented by the green areas. The implementation with externally activated amplitude damping noises follows analogously, with the addition of a polarization rotator set on one spatial mode immediately before the BS. The red wavy lines represent environmentally-induced noise sources.
            The D element represents a pseudospin-insensitive, non-absorbing parity check detector. This setup is applicable to generic bosons and fermions by suitably adapting the represented photonic devices. The figure recalls Fig.\tilde1 of Ref.\tilde\cite{piccolini2023asymptotically}.}
    \label{fig:procedure}
    \end{figure}
    
    The considered protocol is schematically depicted in Fig.\tilde\ref{fig:procedure}.
    Two identical qubits are initially localized on two distinct spatial modes L and R, prepared in an initial state $\ket{\psi_0}$. Two arbitrary environmental induced noises act locally on the two spatial modes for a time $t$. After that, local depolarization is is artificially induced on both qubits, leading to a maximally mixed state. The two particles are now let to impinge on the two input ports of a BS. A pseudospin-insensitive, non-absorbing, parity check detector D is employed on one of the output modes to discriminate the outcomes with one constituent per mode (odd parity) from the ones with two particles in the same spatial mode (even parity). When dealing with bosons, the state is collected in the former case, terminating the process. If an even-parity state is found, instead, the method is iterated by depolarizing the particles and detecting them again, until an odd-parity state is distilled.
    In an alternative implementation, two local amplitude damping channels are employed instead of the depolarizing ones. In this case, a pseudospin rotator, which maps the pseudospin $\uparrow$ of a qubit into $\downarrow$ and viceversa, is set on one spatial mode immediately before the BS. In Ref.\tilde\cite{piccolini2023asymptotically}, the authors showed that this procedure allows to distillate bosonic Bell singlet states, which can be afterward converted into arbitrary maximally entangled states via PO operations (see Subsection\tilde\ref{subsec:fermionicPO}) and, if required, the action of the detector D.

    We stress that, when the externally activated depolarizing (amplitude damping) channels are let to act on the qubits for a sufficiently long time, they lead to the maximally mixed state (ground state) independently on both the initial state $\ket{\psi_0}$, the interaction time $t$, and the type of environmental noise acting before them: in this sense, the artificially induced noises are used to reset the system to a fixed state. On the contrary, all the mentioned factors contribute to the final outcomes when such channels are not activated.

    \section{Fermionic implementation}
    \label{sec:fermionicimpl}
    We introduce the following notation for maximally entangled states of two qubits distributed over two distinct spatial modes L, R:
    \begin{equation}
    \label{maximallystates}
        \begin{gathered}
            \ket{1_\pm}_{\scriptscriptstyle\textrm{LR}}
            :=\frac{1}{\sqrt{2}}\Big(\klurd\pm\kldru\Big),\\
            \ket{2_\pm}_{\scriptscriptstyle\textrm{LR}}
            :=\frac{1}{\sqrt{2}}\Big(\kluru\pm\kldrd\Big),\\
            \ket{1_\pm}_{\scriptscriptstyle\textrm{NO}}
            :=\frac{1}{\sqrt{2}}\Big(\kluld\pm\krurd\Big),
            \\
            \ket{U_\pm}_{\scriptscriptstyle\textrm{NO}}
            :=\frac{1}{2}\Big(\klulu\pm\kruru\Big),
            \\
            \ket{D_\pm}_{\scriptscriptstyle\textrm{NO}}
            :=\frac{1}{2}\Big(\kldld\pm\krdrd\Big).
        \end{gathered}
    \end{equation}
    Here, states with subscripts LR and NO are Bell states and NOON states, respectively. The two differ in the number of particles localized in one spatial mode, which is 1 for the former and 0 or 2 for the latter and can thus be discriminated by a parity measurement. States in Eq.\tilde\eqref{maximallystates} constitute a basis $\mathcal{B}:=\base{LR}\cup\base{NO}$ of the 10-dimensional bosonic Hilbert space, where $\base{LR}:=\Big\{\ket{1_\pm}_{\scriptscriptstyle\textrm{LR}},\,\ket{2_\pm}_{\scriptscriptstyle\textrm{LR}}\Big\}$ and $\base{NO}:=\Big\{\ket{1_\pm}_{\scriptscriptstyle\textrm{NO}},\,\ket{U_\pm}_{\scriptscriptstyle\textrm{NO}},\,\ket{D_\pm}_{\scriptscriptstyle\textrm{NO}}\Big\}$.
    On the other hand, Pauli exclusion principle forbids the existence of states $\ket{U_\pm}_{\scriptscriptstyle\textrm{NO}},\,\ket{D_\pm}_{\scriptscriptstyle\textrm{NO}}$ for identical fermions, restricting their Hilbert space to 6 dimensions spanned by the remaining vectors of $\mathcal{B}$.

        \subsection{Passive optical operations}
        \label{subsec:fermionicPO}
        Referring to a photonic implementation, \textit{PO operations} are defined as the set of transformations which can be obtained by a proper sequence of BSs, polarization BSs (PBSs), polarization-dependent or -independent phase shifters (PDPSs/PIPSs), and local polarization rotators (PRs), with two states being 
        \textit{PO equivalent} if they can be obtained one from the other by means of PO operations\tilde\cite{piccolini2023asymptotically}. Here, we extend PO operations to fermions by simply asking for the involved devices to act on the particle pseudospin rather than polarization, performing analogous transformations.
        
        \begin{figure}[t!]
            \includegraphics[width=0.95\columnwidth]{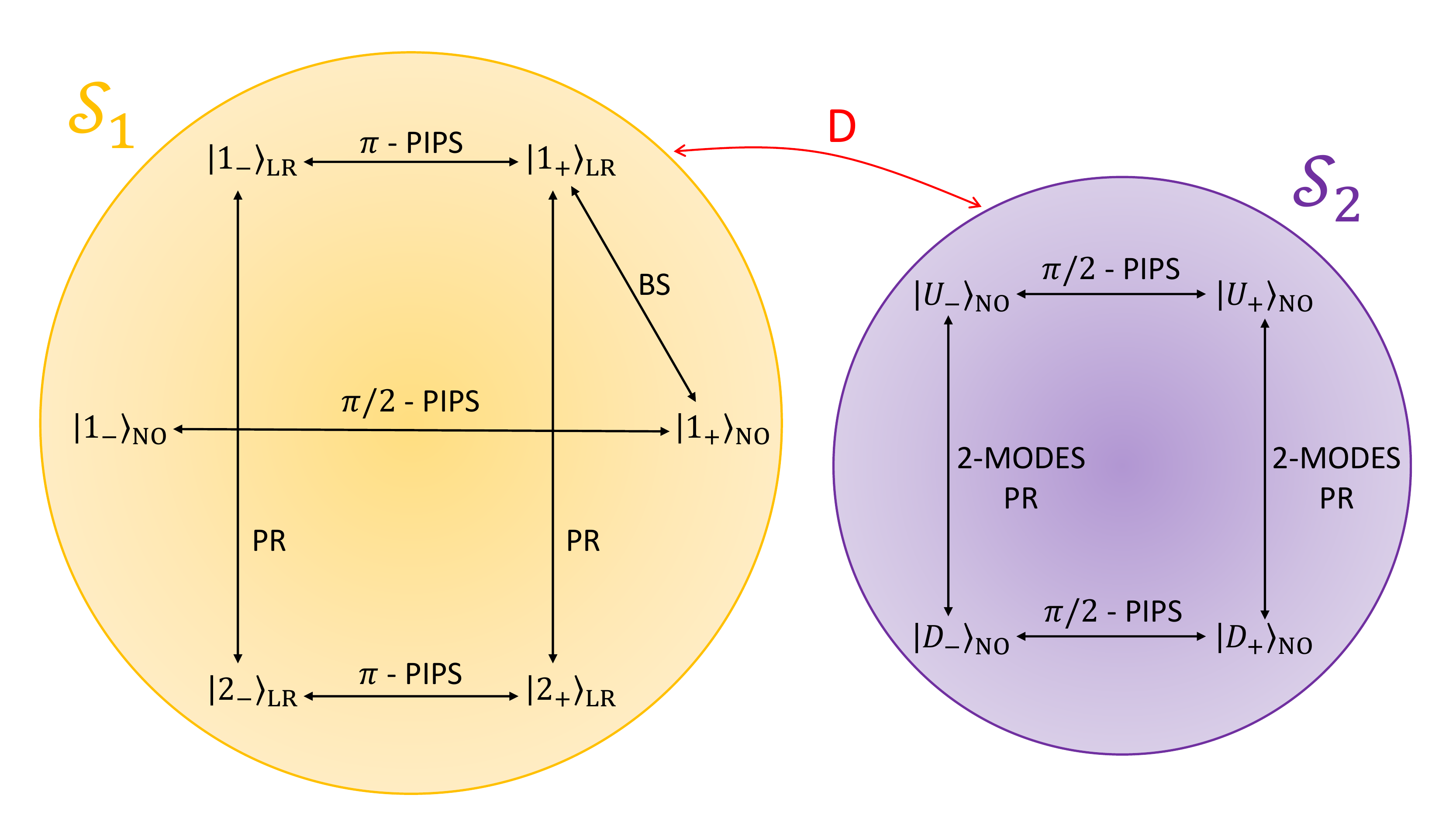}
            \centering
                \caption{\textbf{Structure of passive optical equivalent maximally entangled states of two photons.} The figure shows two sets of PO equivalent maximally entangled states of two bosonic qubits distributed over two spatial modes. Examples of PO transformations connecting them are reported for each set. All the depicted PO transformations are assumed to occur on a single arbitrary spatial mode, except when "2-modes" is stated. $\theta$-PIPS are polarization independent phase shifter inducing a phase $\theta$ on the spatial mode they are set on, PRs are $90^\circ$ polarization rotators, and BSs are beam splitters. The two sets are linked by a polarization-insensitive, non-absorbing, parity check detector D (see main text). The figure recalls Fig.\tilde2 of Ref.\tilde\cite{piccolini2023asymptotically}.}
        \label{fig:bosonicposcheme}
        \end{figure}
        
\begin{figure}[t!]
            \includegraphics[width=0.5\columnwidth]{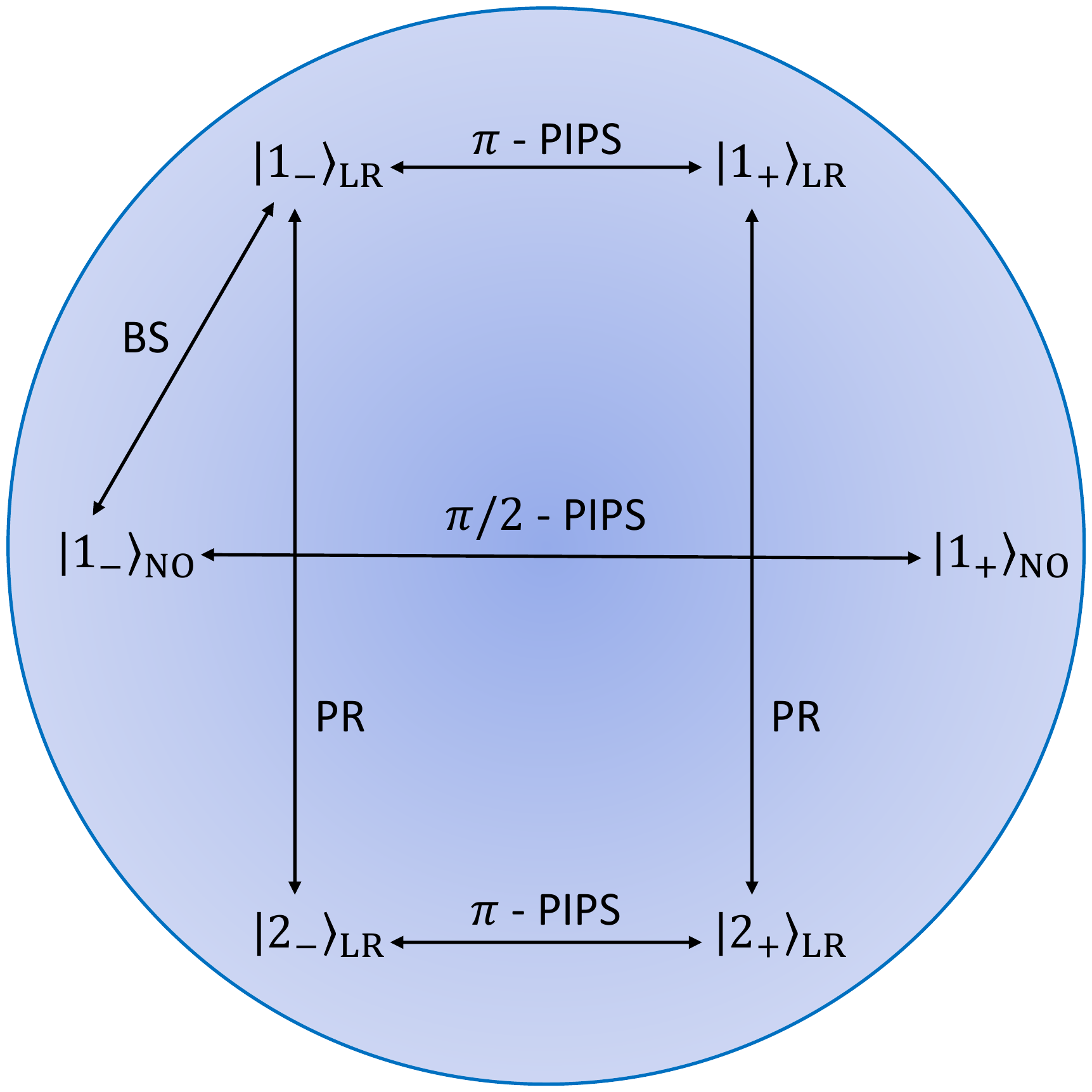}
            \centering
            \caption{\textbf{Structure of passive optical equivalent maximally entangled states of two fermionic qubits.} The figure shows two sets of PO equivalent maximally entangled states of two fermionic qubits distributed over two spatial modes. Examples of PO transformations connecting them are reported. All the depicted PO transformations are assumed to occur on a single arbitrary spatial mode. $\theta$-PIPS, PRs, and BSs are devices acting on fermions performing operations analogous to their bosonic counterpart (see main text and Fig.\tilde\ref{fig:bosonicposcheme}).}
        \label{fig:fermionicposcheme}
        \end{figure}

         Fig.\tilde\ref{fig:bosonicposcheme} reports the two sets of PO equivalent maximally entangled states of two bosons, with an example of PO transformations connecting them (see Ref.\tilde\cite{piccolini2023asymptotically}).
        Similar relations can be found for fermions, as illustrated in Fig.\tilde\ref{fig:fermionicposcheme}. In particular, a PIPS introducing a phase $\pi$ on one spatial mode links $\kom{LR}$ to $\kop{LR}$ and $\ktm{LR}$ to $\ktp{LR}$, while the introduction of a phase $\pi/2$ transforms $\kom{NO}$ and $\kop{NO}$ into each other. A PR mapping $\uparrow$ into $\downarrow$ and viceversa set on one spatial mode achieves the connections $\kom{LR}\leftrightarrow\ktm{LR}$ and $\kop{LR}\leftrightarrow\ktp{LR}$. Being inherently local, this net of PO equivalent states is the same for fermions and bosons (see Ref.\tilde\cite{piccolini2023asymptotically}). On the contrary, a BS transforms two simultaneously impinging fermions differently from the bosonic situation, due to their different commutation/anticommutation rules. As a 50:50 BS transforms single particle states according to $\ket{L}\longrightarrow(\ket{L}+\ket{R})/\sqrt{2}$ and $\ket{R}\longrightarrow(\ket{L}-\ket{R})/\sqrt{2}$, we find its action on fermionic maximally entangled states to achieve the transformations
        \begin{equation}
        \label{bsactionfer}
            \text{fermionic BS:}
            \,
            \left\{
            \begin{aligned}
                &\kom{LR}\longleftrightarrow\kom{NO},\\
                &\kop{LR}\longleftrightarrow-\kop{LR},\\
                &\ktm{LR}\longleftrightarrow-\ktm{LR},\\
                &\ktp{LR}\longleftrightarrow-\ktp{LR},
            \end{aligned}
            \right.
        \end{equation}
        whereas for bosons we have
        \begin{equation}
        \label{bsactionbos}
            \text{bosonic BS:}
            \,
            \left\{
            \begin{aligned}
                &\kom{LR}\longleftrightarrow-\kom{LR},\\
                &\kop{LR}\longleftrightarrow\kom{NO},\\
                &\ktm{LR}\longleftrightarrow(\kum{NO}-\kdm{NO})/{\sqrt{2}},\\
                &\ktp{LR}\longleftrightarrow(\kum{NO}+\kdm{NO})/{\sqrt{2}}.
            \end{aligned}
            \right.
        \end{equation}
        A 50:50 BS can thus be employed to connect the fermionic states $\kom{LR}$ and $\kom{NO}$. Since $\ket{U_\pm}_{\scriptscriptstyle\textrm{NO}}$ and $\ket{D_\pm}_{\scriptscriptstyle\textrm{NO}}$ are forbidden for identical fermions, we thus find that all the fermionic maximally entangled states are PO equivalent. The described situation is pictorially depicted in Fig.\tilde\ref{fig:fermionicposcheme}.
        
        \subsection{Results for fermions}
        \label{subsec:fermionicresults}

        Let us consider the implementation of the protocol based on the artificially induced depolarization of the two qubits. Once the depolarization is complete, the system is left in a maximally mixed state $\rdep:=\frac{1}{4}\,\proj{LR}$, where $\proj{LR}:=\sum_{\ket{v}\in\base{LR}}\ket{v}\bra{v}$ is the projector onto the subspace spanned by the elements of the basis $\base{LR}$ introduced in Section\tilde\ref{sec:fermionicimpl}. Under the action of a BS given in Eq.\tilde\eqref{bsactionfer}, state $\rdep$ is transformed into
        \begin{equation}
        \label{rbsfer}
            \rbs=\frac{1}{4}\kom{NO}\bom{NO}+\frac{3}{4}\rlr,
        \end{equation}
        where
        \begin{equation}
        \label{rlr}
            \rlr:=
            \frac{1}{3}\Big(
            \kop{LR}\bop{LR}+\ktp{LR}\btp{LR}+\ktm{LR}\btm{LR}\Big).
        \end{equation}
        The previously discussed detector D can now be employed to discriminate the even-parity components of $\rbs$ from the odd-parity ones. In the first case, $\rbs$ is projected onto the subspace spanned by the elements of basis $\base{NO}$ via the projection operator $\proj{NO}:=\sum_{\ket{k}\in\base{NO}}\ket{k}\bra{k}$, returning $\kom{NO}\bom{NO}$ with probability $\pno=\trace{\proj{NO}\rbs}=1/4$. We thus collect the qubits, and the process terminates. In the second case, instead, $\rbs$ is projected onto the subspace spanned by the elements of basis $\base{LR}$ via $\proj{LR}$, returning $\rlr$ with probability $\plr=\trace{\proj{LR}\rbs}=3/4$. When this happens, the two qubits are artificially depolarized once again, re-obtaining $\rdep$ and allowing for the process to be iterated.
        % Notice that a closed configuration where the qubits detected in the state $\rlr$ are reflected back into the same BS before being depolarized again is actually possible, since $\rlr$ is left invariant by the action of the BS (see Eq.\tilde\eqref{bsactionfer}).
        By repeating the procedure, we find the probability of distilling the state $\kom{NO}\bom{NO}$ at the j-th iteration to be $\pno^{(j)}=\sum_{n=1}^{j}(1/4)(3/4)^{n-1}$, which converges exponentially to 1 for $j\to\infty$. Given the PO transformations discussed in Subsection\tilde\ref{subsec:fermionicPO}, we thus conclude that the proposed scheme allows for the robust preparation of any maximally entangled state of two fermionic qubits in an asymptotically-deterministic way.

        We now consider the implementation where the artificially induced noise is an amplitude damping locally acting on both particles, followed by a PR set on one spatial mode, which here we fix to be the L one for simplicity. When the damping is complete, the system is left in the ground state $\kldrd$ regardless of the initial state $\ket{\psi_0}$, of the type of noise previously acting on the qubits, and of the interaction time $t$. The PR transforms it into $\klurd=(\kom{LR}+\kop{LR})/\sqrt{2}$, which is then transformed by a BS. From Eq.\tilde\eqref{bsactionfer}, the resulting state is easily found to be $(\kom{NO}-\kop{LR})/\sqrt{2}$. Once again, the detector D allows for discriminating the even component $\kom{NO}$ from the odd one $\kop{LR}$, both distilled with the same probability $\pno=\plr=1/2$. Depending on whether we are interested in preparing a NOON state or a Bell state, we collect the particles when the corresponding parity is detected and repeat the process otherwise. In this way, the desired state is obtained at the j-th iteration with probability $\pno^{(j)}=\plr^{(j)}=\sum_{n=1}^{j} 1/2^{n}$, which for $j\to\infty$ converges to 1 exponentially.
        Nonetheless, differently from the depolarizing implementation, we stress that in this case both even and odd components are pure states, so that the desired state can always be obtained from the distilled one by means of the PO transformations previously discussed. Thus, we conclude that the implementation employing artificial amplitude damping channels enables the robust preparation of any maximally entangled state of two fermionic qubits in a way which is deterministic already in just one run of the protocol.

    \section{Implementation without externally activated noisy channels}
    In this Section, we analyze the situation where the artificial introduction of externally activated noise is not possible. Given its role of resetting the qubits to a determined state, dropping this assumption introduces new factors which ultimately determine the performances of the setup. In particular, the initial state $\ket{\psi_0}$, the type of environmental induced noise acting on the particles before the BS, and the interaction time $t$ all affect the distilled state. In this work, we consider the pure initial state
    \begin{equation}
    \label{init1}
        \ket{\psi_0}=a\klurd+b\,e^{i\phi}\kldru.
    \end{equation}
    % and combinations of parallel pseudospin ones, that is,
    % \begin{equation}
    % \label{init2}
    %     \kpsip=c\kluru+\sqrt{1-c^2}\,e^{i\phi}\kldrd.
    % \end{equation}
    Here, $a$ is a real and positive number, $b:=\sqrt{1-a^2}$, and $\phi\in[0,2\pi)$.
    We highlight that the quantum correlations carried by $\ket{\psi_0}$ depend on the parameter $a$, ranging from separable states for $a=0,1$ to maximally entangled ones for $a=1/\sqrt{2}$. Notice that, in the latter situation, we recover the Bell states $\kop{LR},\,\kom{LR}$ in Eq.\tilde\eqref{maximallystates} when $\phi=0,\pi$, respectively.
    We study three standard models of local noisy environments: phase damping channels, depolarizing channels, and amplitude damping channels. Within this framework, the results discussed in Ref.\tilde\cite{piccolini2023asymptotically} for bosons and in the previous Section for fermions provide the limiting case of the two latter types of noise when $t\to\infty$, with the addition of a PR on one mode in the amplitude damping scenario. Particle statistics is taken into account, to highlight the different performances of bosons and fermions. Furthermore, we focus on a non-iterated version of the process.
    % postponing the study of an asymptotically-deterministic implementation to future works in exchange of a more plain discussion here.

        \subsection{Characterization of the noisy environments}
        We consider the two noisy environments to be of the same type and to independently act on one spatial mode each. We model them as baths of harmonic oscillators at zero temperature, the interaction with the particles being described by a qubit-cavity model where the coupling involves one single excited mode of the baths\tilde\cite{breuer2002theory}. We assume the spectral density of the baths to be Lorentzian, i.e.\tilde\cite{breuer2002theory,Haikka_2010},
        \begin{equation}
        \label{lorentzdensity}
            J(\omega)=
            \frac{\gamma}{2\pi}
            \frac{\lambda^2}{(\omega-\omega_0)^2+\lambda^2},
        \end{equation}
        where $\gamma$ is the coupling strength between the system and the related environment, $\lambda$ is the spectral width of the coupling, and $\omega_0$ is the qubit transition frequency. We compute the dynamics of the system using the operator-sum formalism\tilde\cite{breuer2002theory,nielsen2010quantum}.
        % Denoting with $\rlr$ the two-particle state, its evolution is obtained by the equation
        % \begin{equation}
        % \label{kraus}
        %     \rlr(t)=\sum_{i,j} E_{ij} \rlr(0) E_{ij}^\dagger,
        % \end{equation}
        % where $E_{ij}=E_i^{\text{L}}\otimes E_j^{\text{R}}$ are given by the single particle Kraus operators $E_i^X$ characterizing the environment which acts on the mode $X=$ L, R.
        While the structure of the employed Kraus operators depends on the specific type of environment considered, they all encompass the time-dependent disturbance probability $p(t)$, which for the Lorentzian spectral density in Eq.\tilde\eqref{lorentzdensity} is given by\tilde\cite{breuer2002theory,Bellomo_2007}
        \begin{equation}
        \label{probabilityfunc}
            p(t)
            =1-e^{-\lambda t}\left[\cos\left(\frac{d\,t}{2}\right)+\frac{\lambda}{d}\sin\left(\frac{d\,t}{2}\right)\right]^2,
        \end{equation}
        with $d:=\sqrt{2\gamma\lambda-\lambda^2}$.

        \subsection{Phase damping channel}
        Under the action of two local phase damping channels, the initial state $\rho_0=\ket{\psi_0}\bra{\psi_0}$ evolves into
        \begin{equation}
        \label{rhoaevolvedpd}
            \begin{aligned}
                \rhogen{pd}{}(t)
                &=\Big(1-p(t)\Big)\ket{\psi_0}\bra{\psi_0}\\
                &+ \frac{p(t)}{2}\,\Big(\kop{LR}\bop{LR}+\kom{LR}\bom{LR}\Big)\\
                &+ \frac{p(t)}{2}\,(2a^2-1)\,\Big(\kop{LR}\bom{LR}+\kom{LR}\bop{LR}\Big).
            \end{aligned}
        \end{equation}

        % Similarly, the initial state $\rhogen{p}{}=\kpsip\bpsip$ becomes
        % \begin{equation}
        % \label{rhopevolvedpd}
        %     \begin{aligned}
        %         \rhogen{p}{}(t)
        %         &=\Big(1-p(t)\Big)\kpsip\bpsip\\
        %         &+ p(t)\,\Big(\ktp{LR}\btp{LR}+\ktm{LR}\btm{LR}\Big)\\
        %         &+ p(t)\,(2a^2-1)\,\Big(\ktp{LR}\btm{LR}+\ktm{LR}\btp{LR}\Big).
        %     \end{aligned}
        % \end{equation}

                   \begin{figure}[t!]
                \includegraphics[width=1\columnwidth]{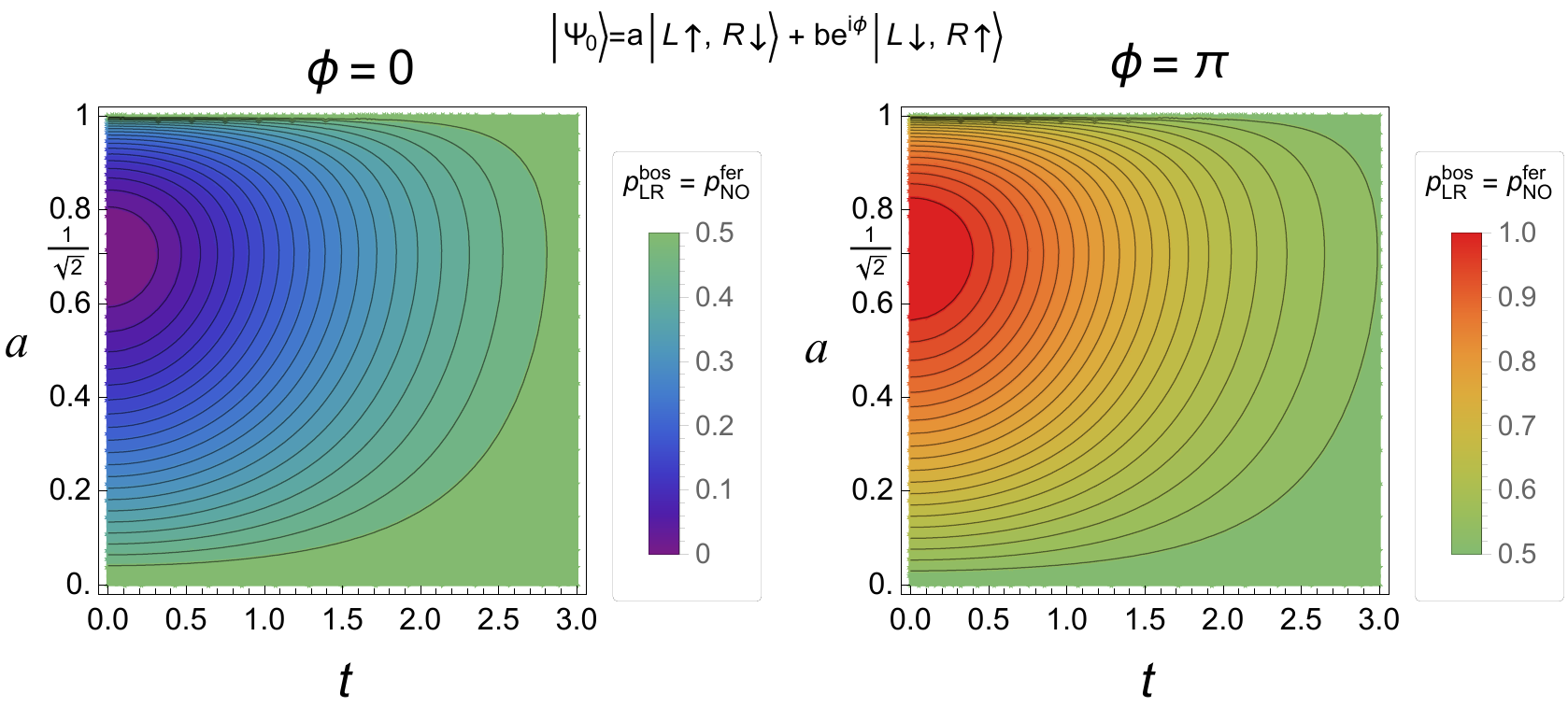}
                \centering
                \caption{\textbf{Distillation probability} of the pure Bell state $\rhogen{LR}{bos}$ (NOON state $\rhogen{NO}{fer}$) of two bosonic (fermionic) qubits subjected to local \textbf{phase damping} for a time $t$. A Lorentzian spectral density is assumed (see Eq.\tilde\eqref{lorentzdensity}), with $\gamma=\lambda=1$ (non-Markovian regime).}
            \label{fig:pdplot}
            \end{figure}
            
              \textbf{Bosons.} Two identical bosons in the state $\rhogen{pd}{}(t)$ given in Eq.\tilde\eqref{rhoaevolvedpd} impinging on the input ports of a 50:50 BS are transformed according to Eq.\tilde\eqref{bsactionbos} into
            \begin{equation}
            \label{rhoabspdbos}
                \begin{aligned}
                    \rhogen{BS}{bos}(t)
                    &=\Big[\frac{1}{2}+\big(1-p(t)\big)\,a\,b\,\cos{\phi}\Big]\kom{NO}\bom{NO}\\
                    &+\Big[\frac{1}{2}-\big(1-p(t)\big)\,a\,b\,\cos{\phi}\Big]\kom{LR}\bom{LR}\\
                    &+\Big[\frac{1}{2}-a^2-i\,\big(1-p(t)\big)\,a\,b\,\sin{\phi}\Big]\kom{NO}\bom{LR}\\
                    &+\Big[\frac{1}{2}-a^2+i\,\big(1-p(t)\big)\,a\,b\,\sin{\phi}\Big]\kom{LR}\bom{NO}.
                \end{aligned}
            \end{equation}
            We thus notice that, by employing the parity check detector D previously introduced, we can distill either the pure NOON state $\rhogen{NO}{bos}:=\kom{NO}\bom{NO}$ with probability $\pgen{NO}{bos}(t)=\frac{1}{2}+\big(1-p(t)\big)\,a\,b\,\cos{\phi}$ or the pure Bell state $\rhogen{LR}{bos}=\kom{LR}\bom{LR}$ with probability $\pgen{LR}{bos}(t)=\frac{1}{2}-\big(1-p(t)\big)\,a\,b\,\cos{\phi}$.
            Fig.\tilde\ref{fig:pdplot} depicts $\pgen{LR}{bos}(t)$ as a function of $a$ and of the interaction time $t$, for $\phi=0,\,\pi$. We notice that, as $t\to\infty$, both $\pgen{LR}{bos}(t)\to1/2$ and $\pgen{NO}{bos}(t)\to1/2$ independently on $a$ and $\phi$, that is, on the initial state. When $\ket{\psi_0}$ is separable ($a=0,\,1$), the detections of $\rhogen{LR}{bos}$ and $\rhogen{NO}{bos}$ are equiprobable, with $\pgen{LR}{bos}=\pgen{NO}{bos}=1/2$ independently on the interaction time. Equiprobability is obtained also for $\phi=\pi/2,\,3\pi/2$, independently on both $a$ and $t$.
            When $\ket{\psi_0}$ is entangled and the relative phase is real, instead, it holds that $\pgen{LR}{bos}(\phi=\pi)\geq\pgen{LR}{bos}(\phi=0)$ for any $a$ and $t$, with the Bell singlet state ($a=1/\sqrt{2},\,\phi=\pi$) maximizing the probability of detecting $\rhogen{LR}{bos}$ at finite times. In particular, $\pgen{LR}{bos}(t)$ increases with the interaction time for $\phi=0$ and decreases when $\phi=\pi$.
            Nonetheless, we stress that both $\rhogen{LR}{bos}$ and $\rhogen{NO}{bos}$ are pure, maximally entangled states, either in the pseudospin or in the polarization, with related detection probabilities satisfying $\pgen{NO}{bos}(t)+\pgen{LR}{bos}(t)=1$. Using the bosonic PO equivalences reported in Fig.\tilde\ref{fig:bosonicposcheme}, both the states can be transformed into any other maximally entangled state. Thus, we conclude that considering two bosons subjected to local phase damping, our method allows for the robust and deterministic preparation of any arbitrary pure and maximally entangled state, without the necessity to iterate the process.

            \textbf{Fermions.}
            When the BS operation is applied to two identical fermionic qubits in the state given in Eq.\tilde\eqref{rhoaevolvedpd}, we get (see Eq.\tilde\eqref{bsactionfer})
            \begin{equation}
            \label{rhoabspdfer}
                \begin{aligned}
                    \rhogen{BS}{fer}(t)
                    &=\Big[\frac{1}{2}+\big(1-p(t)\big)\,a\,b\,\cos{\phi}\Big]\kop{LR}\bop{LR}\\
                    &+\Big[\frac{1}{2}-\big(1-p(t)\big)\,a\,b\,\cos{\phi}\Big]\kom{NO}\bom{NO}\\
                    &+\Big[\frac{1}{2}-a^2-i\,\big(1-p(t)\big)\,a\,b\,\sin{\phi}\Big]\kop{LR}\bom{NO}\\
                    &+\Big[\frac{1}{2}-a^2+i\,\big(1-p(t)\big)\,a\,b\,\sin{\phi}\Big]\kom{NO}\bop{LR}.
                \end{aligned}
            \end{equation}
            As for the bosonic scenario, we notice that the parity check detector D allows to discriminate the even-parity component $\rhogen{NO}{fer}:=\kom{NO}\bom{NO}$ of $\rhogen{BS}{fer}$ from the odd-parity one $\rhogen{LR}{fer}=\kom{LR}\bom{LR}$, detected with respective probabilities $\pgen{NO}{fer}(t)=\frac{1}{2}-\big(1-p(t)\big)\,a\,b\,\cos{\phi}$ and $\pgen{LR}{fer}(t)=\frac{1}{2}+\big(1-p(t)\big)\,a\,b\,\cos{\phi}$.
            Furthermore, as $\pgen{NO}{fer}(t)=\pgen{LR}{bos}(t)$ and $\pgen{LR}{fer}(t)=\pgen{NO}{bos}(t)$, the same considerations drawn for the odd-parity (even-parity) component in the bosonic scenario hold for the even-parity (odd-parity) one for fermions, with $\pgen{NO}{fer}(t)$ being reported in Fig.\tilde\ref{fig:pdplot}, too.
            Given the fermionic PO equivalences introduced in Subsection\tilde\ref{subsec:fermionicPO}, we thus conclude that when applied to two fermions subjected to local phase damping, our method allows for the robust and deterministic preparation of any arbitrary pure and maximally entangled state, without having to iterate the process.

        \subsection{Depolarizing channel}
        Let us now consider the two qubits prepared in the initial state $\rho_0=\ket{\psi_0}\bra{\psi_0}$ to be subjected to the local action of a depolarizing channel each.
        Such an interaction leads to the Werner state
        \begin{equation}
        \label{rhoaevolveddep}
            \rhogen{dep}{}(t)
            =\Big(1-p(t)\Big)\ket{\psi_0}\bra{\psi_0}
            +\frac{1}{4}\,p(t)\,\proj{LR},
        \end{equation}
        with $\proj{LR}$ defined in Subsection\tilde\ref{subsec:fermionicresults}.
        As previously mentioned, the limit $t\to\infty$ provides the maximally mixed state studied in Ref.\tilde\cite{piccolini2023asymptotically} (for bosons) and in Subsection\tilde\ref{subsec:fermionicresults} (for fermions).

            \begin{figure}[t!]
                \includegraphics[width=1\columnwidth]{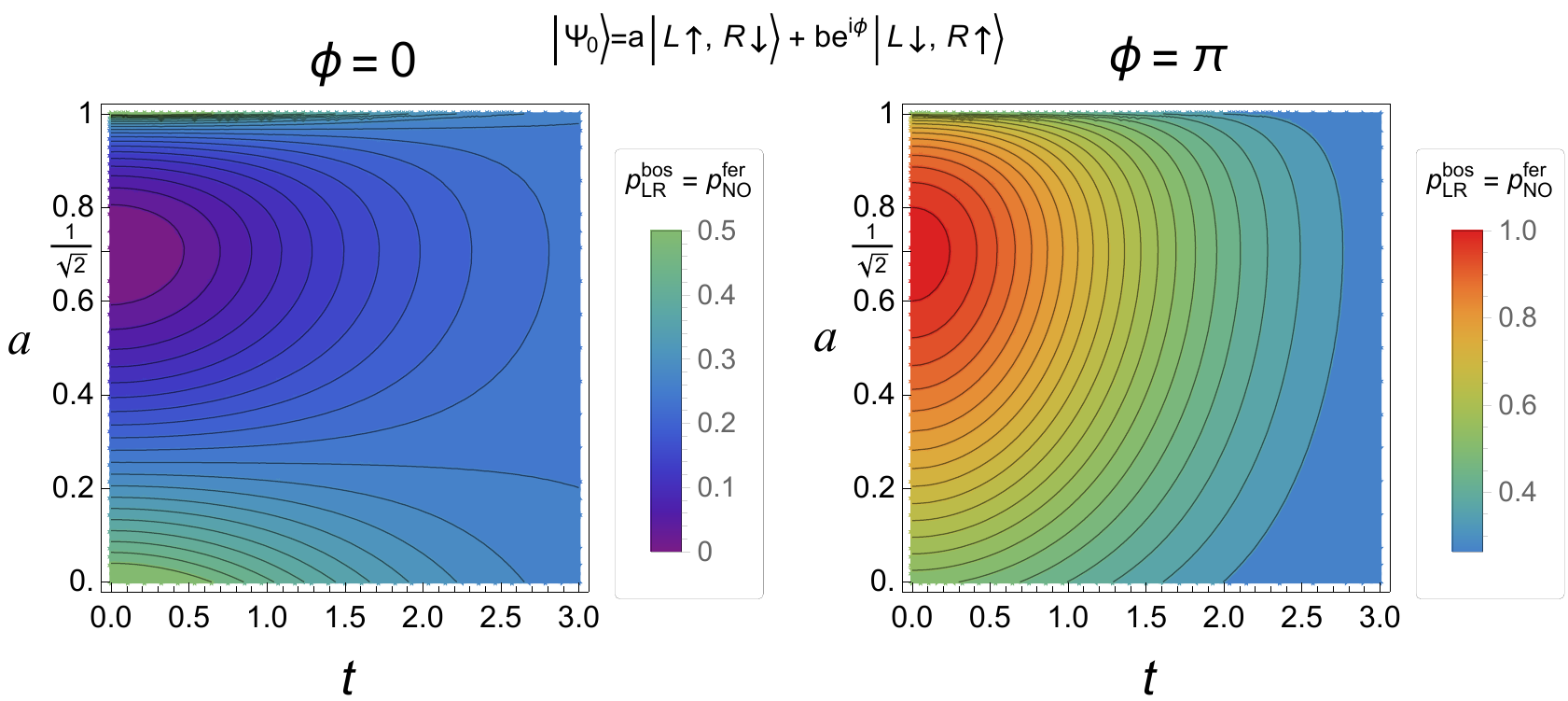}
                \centering
                \caption{\textbf{Distillation probability} of the pure Bell state $\rhogen{LR}{bos}$ (NOON state $\rhogen{NO}{fer}$) of two bosonic (fermionic) qubits subjected to local \textbf{depolarization} for a time $t$. A Lorentzian spectral density is assumed (see Eq.\tilde\eqref{lorentzdensity}), with $\gamma=\lambda=1$ (non-Markovian regime).}
            \label{fig:depplot}
            \end{figure}
            
                \textbf{Bosons.} Two identical bosons in the state $\rdep(t)$ given in Eq.\tilde\eqref{rhoaevolveddep} subjected to the BS operation in Eq.\tilde\eqref{bsactionbos} are transformed into
            \begin{equation}
            \label{rhoabsdepbos}
                \begin{aligned}
                    \rhogen{BS}{bos}(t)
                    &=\Big[\frac{1}{2}-\frac{p(t)}{4}+\big(1-p(t)\big)\,a\,b\,\cos{\phi}\Big]\kom{NO}\bom{NO}\\
                    &+\Big[\frac{1}{2}-\frac{p(t)}{4}-\big(1-p(t)\big)\,a\,b\,\cos{\phi}\Big]\kom{LR}\bom{LR}\\
                    &+\Big(p(t)-1\Big)
                    \big(a^2-\frac{1}{2}+i\,a\,b\,\sin{\phi}\big)\kom{NO}\bom{LR}\\
                    &+\Big(p(t)-1\Big)\big(a^2-\frac{1}{2}-i\,a\,b\,\sin{\phi}\big)\kom{LR}\bom{NO}\\
                    &+\frac{p(t)}{4}\big(\kum{NO}\bum{NO}+\kdm{NO}\bdm{NO}\big).
                \end{aligned}
            \end{equation}
            Once again, the parity check detector D can be employed to distill the pure Bell singlet state $\rhogen{LR}{bos}=\kom{LR}\bom{LR}$ with a probability $\pgen{LR}{bos}(t)=\frac{1}{2}-\frac{p(t)}{4}-\big(1-p(t)\big)\,a\,b\,\cos{\phi}$. $\pgen{LR}{bos}(t)$ is depicted in Fig.\tilde\ref{fig:depplot} as a function of $a$ and $t$ for $\phi=0,\,\pi$.
 
            First of all, $\pgen{LR}{bos}(t)\to 1/4$ as $t\to\infty$ independently on $a$ and $\phi$, i.e., on the initial state $\ket{\psi_0}$, in agreement with the results reported in Ref.\tilde\cite{piccolini2023asymptotically} when depolarization is used to reset the system to a maximally mixed state.
            Furthermore, it holds once again that $\pgen{LR}{bos}(\phi=\pi)\geq\pgen{LR}{bos}(\phi=0)$ for any $a$ and $t$. In particular, the Bell singlet state ($a=1/\sqrt{2},\,\phi=\pi$) is found to maximixe $\pgen{LR}{bos}(t)$ at finite times.
            Nonetheless, differently from the phase damping scenario, the interaction time affects the probability of distilling $\rhogen{LR}{bos}$ even when the initial state is separable ($a=0,\,1$), ranging from $\pgen{LR}{bos}=1/2$ to $\pgen{LR}{bos}=1/4$ as $t\to\infty$. The same behaviour is obtained when $\phi=\pi/2,\,3\pi/2$, regardless of $a$.

            \textbf{Fermions.}
            When the two particles impinging on the BS are fermions, the state $\rdep(t)$ in Eq.\tilde\eqref{rhoaevolveddep} is mapped into
            \begin{equation}
            \label{rhoabsdepfer}
                \begin{aligned}
                    \rhogen{BS}{fer}(t)
                    &=\Big[\frac{1}{2}-\frac{p(t)}{4}+\big(1-p(t)\big)\,a\,b\,\cos{\phi}\Big]\kop{LR}\bop{LR}\\
                    &+\Big[\frac{1}{2}-\frac{p(t)}{4}-\big(1-p(t)\big)\,a\,b\,\cos{\phi}\Big]\kom{NO}\bom{NO}\\
                    &+\Big(p(t)-1\Big)
                    \big(a^2-\frac{1}{2}+i\,a\,b\,\sin{\phi}\big)\kop{LR}\bom{NO}\\
                    &+\Big(p(t)-1\Big)\big(a^2-\frac{1}{2}-i\,a\,b\,\sin{\phi}\big)\kom{NO}\bop{LR}\\
                    &+\frac{p(t)}{4}\big(\ktp{LR}\btp{LR}+\ktm{LR}\btm{LR}\big).
                \end{aligned}
            \end{equation}
            Analogously to the bosonic scenario and in agreement with the results discussed in Subsection\tilde\ref{subsec:fermionicresults}, the parity check detector D can be used to distill the NOON state $\rhogen{NO}{fer}=\kom{NO}\bom{NO}$ with probability $\pgen{NO}{fer}(t)=\frac{1}{2}-\frac{p(t)}{4}-\big(1-p(t)\big)\,a\,b\,\cos{\phi}$. Since $\pgen{NO}{fer}(t)=\pgen{LR}{bos}(t)$, Fig.\tilde\ref{fig:depplot} also reports $\pgen{NO}{fer}(t)$ and the same considerations discussed for the distillation of the bosonic Bell singlet state can be drawn for the fermionic NOON state $\kom{NO}$.

        \subsection{Amplitude damping channel}
        Finally, we study the performance of the proposed method when the two qubits are independently subjected to a local amplitude damping channel. In this situation, the initial state $\rho_0=\ket{\psi_0}\bra{\psi_0}$ evolves into
        \begin{equation}
        \label{rhoaevolvedad}
            \rhogen{ad}{}(t)=\Big(1-p(t)\Big)\ket{\psi_0}\bra{\psi_0}
            +p(t)\kldrd\bldrd,
        \end{equation}
        where $\kldrd=(\ktp{LR}-\ktm{LR})/\sqrt{2}$ is the two-particle ground state.
        We highlight that, differently from the analogous scenario in Ref.\tilde\cite{piccolini2023asymptotically} and in Section\tilde\ref{sec:fermionicimpl}, here we do not assume the presence of a PR rotating the pseudospin of particles on one spatial mode before the BS operation.  

\begin{figure}[b!]
                \includegraphics[width=1\columnwidth]{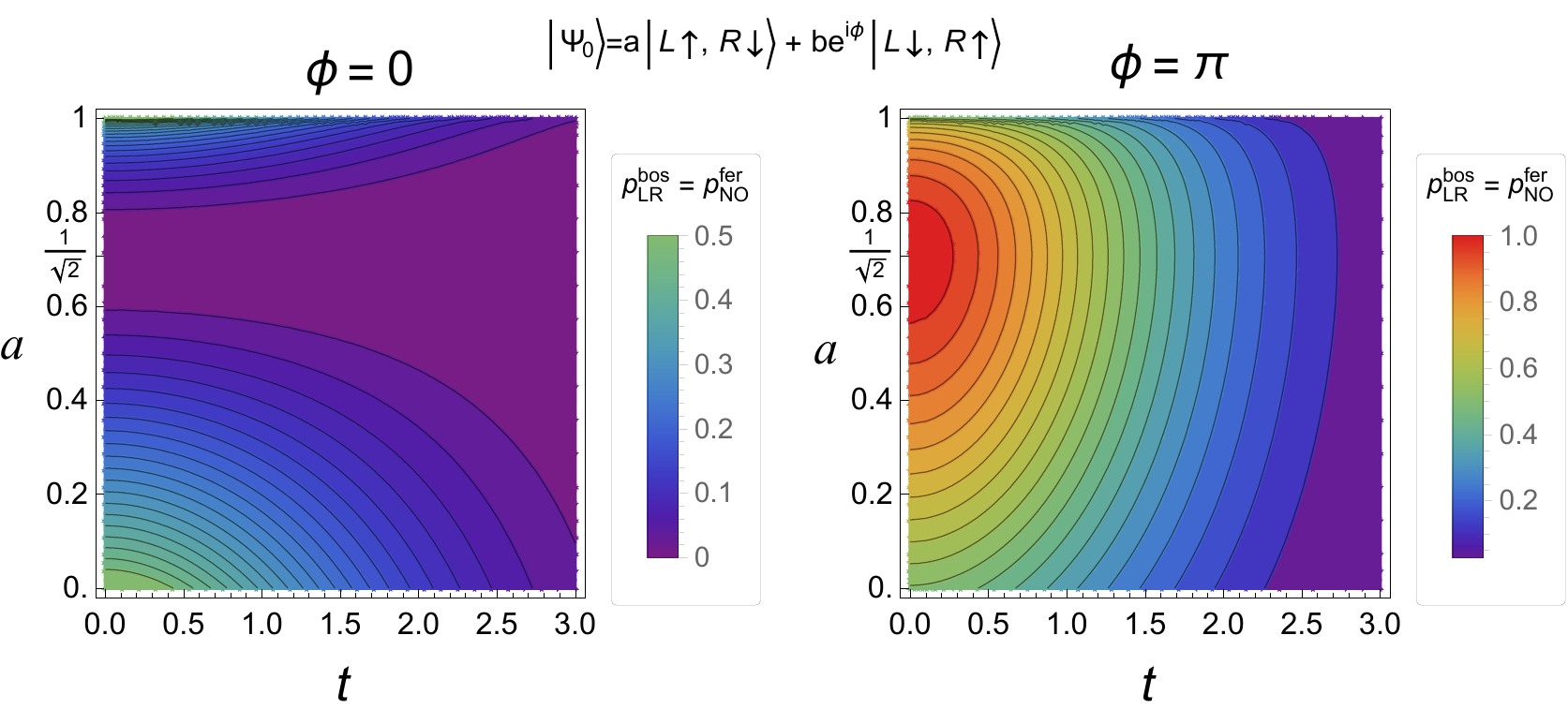}
                \centering
                    \caption{\textbf{Distillation probability} of the pure Bell state $\rhogen{LR}{bos}$ (NOON state $\rhogen{NO}{fer}$) of two bosonic (fermionic) qubits subjected to local \textbf{amplitude damping} for a time $t$. A Lorentzian spectral density is assumed (see Eq.\tilde\eqref{lorentzdensity}), with $\gamma=\lambda=1$ (non-Markovian regime).}
            \label{fig:adplot}
            \end{figure}

            \textbf{Bosons.} The action of a BS on the bosonic state $\rhogen{ad}{}(t)$ given in Eq.\tilde\eqref{rhoaevolvedad} returns
            \begin{equation}
            \label{rhoabsadbos}
                \begin{aligned}
                    \rhogen{BS}{bos}(t)
                    &=\Big(1-p(t)\Big)
                    \bigg[
                        \Big(\frac{1}{2}+a\,b\,\cos{\phi}\Big)\kom{NO}\bom{NO}\\
                        &+\Big(\frac{1}{2}-a\,b\,\cos{\phi}\Big)\kom{LR}\bom{LR}\\
                        &-\Big(a^2-\frac{1}{2}+i\,a\,b\,\sin{\phi}\Big)\kom{NO}\bom{LR}\\
                        &-\Big(a^2-\frac{1}{2}-i\,a\,b\,\sin{\phi}\Big)\kom{LR}\bom{NO}
                    \bigg]\\
                    &+p(t)\kdm{NO}\bdm{NO}.
                \end{aligned}
            \end{equation}
            The detector D allows to distill the pure Bell singlet state $\rhogen{LR}{bos}=\kom{LR}\bom{LR}$ with probability $\pgen{LR}{bos}(t)=(1-p(t))(\frac{1}{2}-a\,b\,\cos{\phi})$, depicted in Fig.\tilde\ref{fig:adplot}.

            As for the phase damping and the depolarizing channels, we notice that $\pgen{LR}{bos}(\phi=\pi)\geq\pgen{LR}{bos}(\phi=0)$ for every $a,\,t$. In particular, such a probability reaches its maximum at finite times when $\ket{\psi_0}$ is a Bell singlet state. However, this time $\pgen{LR}{bos}(t)$ decays to 0 with the interaction time regardless of $a$ and $\phi$. As in this case $\rhogen{NO}{bos}(t)$ is not a pure state (similarly to the depolarizing scenario and in contrast with the phase damping one), at first sight such a decay could be interpreted as an inefficacy of the proposed technique to distill pure and maximally entangled states when the particles of interest have been subjected to the considered noise for too long. Nonetheless, we highlight that in the limit $t\to\infty$ the even parity state $\rhogen{NO}{bos}$ becomes the pure NOON state $\kdm{NO}$, and $\pgen{NO}{bos}(t)=1-\pgen{LR}{bos}(t)\to1$. Thus, for sufficiently long interaction times we should collect the qubits when the detector D signals an even number of particles, achieving the distillation of a state which becomes pure as $t\to\infty$.
            % Given the PO transformations previously discussed, we conclude that the proposed protocol can be employed to distill pure and maximally entangled states of two bosonic qubits in a way which is probabilistic at finite times and deterministic as the interaction time grows to infinity.

            \textbf{Fermions.}
            When dealing with identical fermions, the state $\rhogen{ad}{}(t)$ in Eq.\tilde\eqref{rhoaevolveddep} is transformed by a BS into
            \begin{equation}
            \label{rhoabsadfer}
                \begin{aligned}
                    \rhogen{BS}{fer}(t)
                    &=\Big(1-p(t)\Big)
                    \bigg[
                        \Big(\frac{1}{2}+a\,b\,\cos{\phi}\Big)\kop{LR}\bop{LR}\\
                        &+\Big(\frac{1}{2}-a\,b\,\cos{\phi}\Big)\kom{NO}\bom{NO}\\
                        &-\Big(a^2-\frac{1}{2}+i\,a\,b\,\sin{\phi}\Big)\kop{LR}\bom{NO}\\
                        &-\Big(a^2-\frac{1}{2}-i\,a\,b\,\sin{\phi}\Big)\kom{NO}\bop{LR}
                    \bigg]\\
                    &+p(t)\kldrd\bldrd.
                \end{aligned}
            \end{equation}
            As in the other fermionic scenarios discussed, the detector D allows to distill the pure NOON state $\rhogen{NO}{fer}=\kom{NO}\bom{NO}$. The probability of such a detection is given by $\pgen{NO}{fer}(t)=(1-p(t))(\frac{1}{2}-a\,b\,\cos{\phi})$, which is reported in Fig.\tilde\ref{fig:adplot} as $\pgen{NO}{fer}(t)=\pgen{LR}{bos}(t)$. The considerations drawn for detection of the odd component in the bosonic situation are thus valid also for the detection of $\rhogen{NO}{fer}$. Nonetheless, an important difference emerges: in the limit $t\to\infty$ the fermionic odd-component, which is the only one distillable as $\pgen{LR}{fer}(t)=\pgen{NO}{bos}(t)\to1$, is the pure state $\rhogen{LR}{fer}=\kldrd\bldrd$ which is not entangled. Notice that the presence of this term in $\rhogen{BS}{fer}$ in contrast to $\rhogen{BS}{bos}$ is due to the anti-bunching effect characterizing identical fermions, which makes $\kldrd$ in $\rhogen{ad}{}(t)$ invariant under the action of a BS whereas the bosonic bunching effect transforms it into the maximally entangled NOON state $\kdm{NO}$. Nonetheless, the fermionic state $\kldrd$ can be transformed into a maximally entangled state by means of PO transformations and the detector D as discussed in Subsection\tilde\ref{subsec:fermionicresults}, that is, by rotating the pseudospin of the qubit on one region, by applying a BS operation to the resulting state, and by finally distilling either the odd component $\kop{LR}$ or the even one $\kom{NO}$.

    \section{Conclusions}
    We have extended a theoretical procedure for the robust preparation of maximally entangled states introduced in Ref.\tilde\cite{piccolini2023asymptotically} for bosons to the fermionic scenario.
    Considering two qubits set on two distinct spatial modes and subjected to local environmental noise, the protocol employs two externally activated local noisy channels to reset the bipartite state to a known one.
    % which is either a maximally mixed state when employing depolarizing noise or the ground state when relying on amplitude damping.
    A BS is then used to set the system in a mixture or a superposition of NOON states, entangled in the spatial mode, and Bell states, entangled in the internal degree of freedom.
    % In the former case, both the qubits are both localized in either one spatial mode or the other, being maximally entangled in space; in the latter scenario, instead, each particle is localized in a distinct spatial mode, displaying maximum entanglement in the pseudospin.
    A non-absorbing, pseudospin-independent, parity check detector D set on one mode is later used to discriminate between the two components, distilling a maximally entangled state which is shown to be pure. Such a result is independent on the initial state of the system, on the characteristics of the environments, and on the interaction between the two besides requiring the latter to be local and particle-preserving, making the proposed preparation robust.
    We have analyzed two possible implementations of the protocol: one employing depolarizing channels as the externally activated noises, and one using amplitude damping channels as such. In both cases, the procedure can be iterated; by doing so, the implementation relying on externally activated depolarizing channels achieves determinism asymptotically with the number of iterations, while determinism can be reached either asymptotically or in a single run when employing externally activated amplitude damping channels.
    
    We have extended the bosonic PO equivalences introduced in Ref.\tilde\cite{piccolini2023asymptotically} to fermions, showing that all the pure maximally entangled states of two fermionic qubits distributed on two spatial modes can be obtained one from another by means of transformations analogous to passive optical ones. Provided with such a set of transformations, the proposed framework can thus be employed to deterministically prepare any arbitrary maximally entangled state of two identical fermionic qubits.

    The difficulty in realizing a polarization-insensitive, non-absorbing, parity check detector constitutes the main obstacle hindering an experimental implementation of the proposed scheme. Nonetheless, we emphasize that two commercially available single photon detectors can be used in its place to postselect the odd components by performing a coincidence measurement on the two spatial modes. In this way, the implementation employing amplitude damping channels can be used to achieve the probabilistic distillation of a pure fermionic Bell state with probability $\plr=1/2$.

    We have further studied the performances of our technique when the artificially induced noisy channels are not employed, analyzing both bosonic and fermionic qubits. In this case, the initial state, the characteristics of the environmental noise, and the system-environment interaction time all affect the result. We have taken the two particles to be initially prepared in a generic superposition of anti-parallel pseudospin states as given by Eq.\tilde\eqref{init1}. Such a set of states includes both entangled and separable ones, allowing the role of the initial quantum correlations to be assessed. The two environments acting locally on the qubits have been modeled as identical phase damping channels, depolarizing channels, and amplitude damping channels. Focusing on a single run of the protocol, we have shown that it allows to distillate pure and maximally entangled states deterministically when dealing with phase damping channels. When considering depolarizing channels, instead, a pure maximally entangled state can be prepared conditionally, with a distillation probability decaying to $1/4$ asymptotically with the interaction time regardless of the initial state. Nonetheless, preparing the system in a Bell singlet state increases the chances of success at finite interaction times. Finally, a pure maximally entangled state is probabilistically achieved when the two constituents are affected by amplitude damping noise for a finite time $t$, whereas a deterministic distillation is achievable as $t\to\infty$ regardless of the initial state. Remarkably, all the above results hold for both bosons and fermions.

    We stress that the core of the proposed procedure lies in the bunching/antibunching interferometric effects characterizing identical particles under the action of a BS. In particular, we notice from Eqs.\tilde(\ref{bsactionfer}) and (\ref{bsactionbos}) that the singlet is the only Bell state whose number of particles on one spatial mode changes parity (for fermions) or preserves it (for bosons) under a BS transformation.
    Thus, the parity-check detector ultimately discriminates the pure BS-transformed singlet component from the other ones, achieving the desired distillation. With a specific focus on coincidence measurements, i.e., on the detection of odd-parity terms, this mechanism is also at the heart of the technique employing the \textit{spatially localized operations and classical communication} (sLOCC) operational framework\tilde\cite{slocc,piccolini2022philtrans,experimentalslocc,Barros:20,sun2022activation,wang2022remote,wang2022proof} to recover the quantum correlations initially present in a Bell singlet state subjected to the local action of noisy environments\tilde\cite{indistentanglprotection,Piccolini_2021_entropy,piccolini2021opensys}.
    
    Finally, we emphasize that in order that the discussed interference effects properly occur, the different 2-particle probability amplitudes must be indistinguishable when the qubits are collected. Crucially, as demonstrated in Ref.\tilde\cite{pittman1996can}, this requirement is not  equivalent to ask for the two particles to simultaneously impinge on the BS; an arbitrary time delay is indeed acceptable, provided it gets compensated after the BS. If a time delay occurs due, e.g., to the particle sources being desynchronized, the results of this paper can thus be restored, for example, by suitably arranging the lengths of the different paths after the BS or, referring to a photonic implementation, by inducing a proper dephasing on the particles with birefringent media prior to their collection\tilde\cite{pittman1996can,siltanen2021engineering}.

\textbf{Acknowledgments.} R.L.F. acknowledges support from European Union -- NextGenerationEU -- grant MUR D.M. 737/2021 -- research project ``IRISQ''.
            V.G. acknowledges financial support by MUR (Ministero dell'Universit\`{a} e della Ricerca) through the following projects: PNRR MUR project PE0000023-NQSTI, PRIN 2017 Taming complexity via Quantum Strategies: a Hybrid Integrated Photonic approach (QUSHIP) Id. 2017SRN-BRK, and project PRO3 Quantum Pathfinder.\\

% \section*{Conflict of Interest} The authors declare no conflict of interest.
% 
%\section*{Data Availability Statement} All data generated or analyzed during this study are included in this article.
% 
%\section*{Keywords} identical particles, fermions, bosons, open quantum systems, quantum interferometry, entanglement production, entanglement protection. 

    %\bibliography{Bibliografia}
    

\end{document}